# Close contact risk assessment for SARS-CoV-2 infection

G. Cortellessa[1], L. Stabile[1], F. Arpino[1], D.E. Faleiros[2], W. van den Bos[2], L. Morawska[3], G. Buonanno[1,3]

[1] Department of Civil and Mechanical Engineering, University of Cassino and Southern Lazio, Cassino, FR, Italy
[2] Maritime and Transport Technology, TU Delft, Netherlands
[3] International Laboratory for Air Quality and Health, Queensland University of Technology, Brisbane, Qld, Australia

**\*Corresponding author:**
Giorgio Buonanno, PhD
Full Professor
Department of Civil and Mechanical Engineering,
University of Cassino and Southern Lazio, Cassino, FR, Italy
e-mail: buonanno@unicas.it

**Abstract**

Although close contact represents an important contagion route, the mechanism of exposure to exhaled droplets remains insufficiently characterized. In this study, an integrated risk assessment is presented for SARS-CoV-2 close contact exposure between a speaking infectious subject and a susceptible subject. It is based on a three-dimensional transient numerical model for the description of exhaled droplet spread once emitted by a speaking person, coupled with a recently proposed SARS-CoV-2 emission approach. Particle image velocimetry measurements were conducted to validate the numerical model.
The contribution of large droplets to infection risk is dominant for distances < 0.6 m, whereas for longer distances, the exposure risk depends only on airborne droplets. In particular, for short exposures (10 s) a minimum safety distance of 0.75 m should be maintained to lower the risk below 0.1%; for exposures of 1 and 15 min this distance increases to about 1.0 and 1.5 m, respectively. Based on the interpersonal distances across countries reported as a function of interacting individuals, cultural differences, and environmental and sociopsychological factors, the approach presented here revealed that, in addition to intimate and personal distances, particular attention must be paid to exposures longer than 1 min within social distances (of about 1 m).

**Keywords**: CFD analysis; virus transmission; close contact; PIV; SARS-CoV-2; droplets

## 1. Introduction

The COVID-19 pandemic has highlighted the key role of droplets exhaled during respiratory activities (such as oral breathing, speaking, sneezing, coughing) as potential virus carriers leading to risk of infections and/or diseases (Morawska and Cao, 2020). Three possible routes of transmission are generally considered: the fomite, and large droplet and airborne droplet routes (Li, 2021a).

The large droplet route has been incorrectly considered to be the primary route for most respiratory infections since the beginning of the last century (Chapin, 1912; Flügge, 1897), and the associated protective measure of social distancing of 1–2 m (varying in each country as a function of activities) is known and imposed worldwide. The large droplet route is commonly identified with close contact during which an infectious bacterium or virus can be transmitted effectively via specific routes, as discussed below. Despite the importance, a recent search of the literature revealed that the characterization of exposure to droplets exhaled at close contact remains surprisingly unexplored. We point out that the term of "droplets" in this paper refers to all sizes, from the smallest droplets of sub-micron size to the largest ones with dimensions of hundreds of microns. Recent studies have identified two major routes for close contact transmission (Chen et al., 2020): the large droplet route and the airborne droplet route. The large droplet route concerns the deposition of large droplets from the infected subject on the mucous membranes (lips, eyes, nostrils) of the susceptible subject (Gallo et al., 2021; Lu et al., 2020). According to Xie et al. (2007), who revisited previous guidelines by Wells (1934) with improved evaporation and settling models, droplets larger than about 100 μm in diameter rapidly settle out of the air by gravity, with the infective range being within a short distance of the source. In contrast, the airborne droplet route involves droplets smaller than about 100 μm in diameter that, as soon they are exhaled, decrease their diameter by evaporation and can be inhaled by the receiving host as interconnected multiphase flow processes (Balachandar et al., 2020). In fact, the surrounding ambient air enables the exhaled droplets to evaporate and rapidly shrink to droplet nuclei, reaching a final size that is governed by the initial amount of non-volatiles. As an example, by considering the average peak value of exhaled air velocity for breathing and speaking and the evaporation time, a droplet with a 50 μm diameter would shrink to a droplet nucleus before it reached a nearby person (Zhang et al., 2020a).

Recently Chen et al. (2020), using a simple mathematical model of exhaled flows and both droplet deposition and inhalation phenomena, found the airborne droplet route to dominate in the social distancing range of 1–2 m during both speaking and coughing. The large droplet route (> 100 μm) only dominates when the distance is lower than 0.2 m while talking or 0.5 m while coughing, whereas when the subjects are more than 0.3 m apart, the large droplet route can be neglected even while coughing.

Although the work of Chen et al. (2020) represents a novel analysis in the investigation of infection transmission in close contact, it has limitations mainly associated with the simplicity of the proposed analytical model, i.e. adopting steady state conditions, not considering the fluid dynamics connected to the breathing of the subjects, and using corrective coefficients to simulate the inhalation process. In this regard, thermo-fluid dynamic modeling represents a more advanced approach to solve complex flow behaviors (i.e. three-dimensional, transient) typical of respiratory activities, and of the related droplet emission and inhalation, which are not possible with ordinary calculus. However, to date, most of the thermo-fluid dynamic analyses performed on this very complex subject are limited to non-pathogen carrying droplet emission and/or simulate simplified conditions (e.g. constant emission, steady-state analyses) (Ai et al., 2019; Ai and Melikov, 2018).

A further limitation of the analysis proposed by Chen et al. (2020) is that it estimates the fluid dynamics of droplets but does not investigate the issues related to host-to-host viral airborne transmission, which is fundamental in an infectious risk assessment. To this end, the authors recently presented an approach to evaluate the viral load emitted by infected individuals (Buonanno et al., 2020b) that takes into account the effect of other parameters such as inhalation rate, type of respiratory activity, and activity level. Such an approach has been applied to prospective and retrospective cases (Buonanno et al., 2020a), and then extended to viruses other than SARS-CoV-2 (Mikszewski et al., 2021), adopting a simplified zero-dimensional model and allowing the risk of infection due to airborne droplets to be estimated in different indoor scenarios involving people sharing room air and maintaining distancing. Nonetheless, the well-mixed hypothesis of such an approach cannot be applied when it comes to evaluating the risk of infection in close contact scenarios.

On the basis of the abovementioned approach allowing evaluation of the viral load emitted, in this paper the authors present an integrated approach aimed at assessing the close contact risk of infection from SARS-CoV-2. For this purpose, a numerical approach has been developed to estimate the volume of the droplets and the corresponding viral load received by a susceptible subject (through inhalation and deposition) at different distances in close contact scenarios (distance less than 2 m). Particle image velocimetry (PIV) measurements were conducted to characterize the air flow exhaled during human expiratory activities to validate the modelling results. Therefore, on the basis of the integrated approach between thermo-fluid dynamic modeling of exhaled droplets and viral load, an infectious risk assessment is presented for a close contact scenario represented by a speaking infected subject (emitter) and a susceptible subject (receiver) in the case of a face-to-face orientation and stagnant air conditions.

## 2. Materials and methods

The proposed SARS-CoV-2 infectious risk assessment is characterized by an integrated approach, based on the following main steps: (i) development of a three-dimensional Eulerian-Lagrangian numerical model to describe droplet spread once emitted by a speaking person in transient conditions; this is based on an Eulerian-Lagrangian approach, in which the continuum equations are solved for the air flow and Newton's equation of motion is solved for each droplet (sections 2.1, 2.2); (ii) PIV measurements to define the boundary conditions and to validate the numerical model (section 2.3); iii) definition of a droplet emission model including droplet diameters from 0.5 μm to 800 μm emitted by an adult while speaking (section 2.4); and iv) infectious SARS-CoV-2 risk assessment in a close contact scenario by considering the contributions of the large droplet and airborne droplet routes as well as the distance between the speaking infected subject and a susceptible subject (section 2.5). The assessment has been performed in stagnant air conditions, which clearly represents the worst scenario in terms of virus spread as it could occur also in outdoor environments with negligible wind speed.

*2.1. Eulerian-Lagrangian based model to simulate the droplet dynamics at close contact*

The Computational Fluid Dynamics (CFD) technique has been adopted for numerical description of velocity, pressure, and temperature fields, along with the motion and interaction of the droplets with the fluid. The fully open source finite volume based openFOAM software has been employed. This choice was dictated by the need to have a fully open and flexible tool with complete control over the solved partial differential equations (PDEs), boundary conditions, and correlations employed for SARS-CoV-2 risk assessment. The substantial complexity of the adopted approach has paid off by allowing detailed description in space and time of thermo-fluid dynamic fields and associated droplet motion. Additionally, the use of the openFoam software offered the ability to directly access the source code, so modifying available mathematical models, boundary conditions, and thermophysical models is possible, as well as implementing new ones if necessary.

From a mathematical point of view, the droplet motion inside the air flow has been modeled by employing the Lagrangian particle tracking (LPT) approach, based on a dispersed dilute two-phase flow. In particular, the spacing between droplets in the exhaled air plume is sufficiently large and the volume fraction of the droplets sufficiently low ($< 10^{-3}$) to justify the use of a Eulerian-Lagrangian approach, in which the continuum equations are solved for the air flow (continuous phase) and Newton's equation of motion is solved for each droplet. The continuum equations solved for an unsteady incompressible Newtonian fluid are widely described in the available scientific literature (Arpino et al., 2014; Massarotti et al., 2006; Scungio et al., 2013) and are not reported here for brevity. Since the flow regime associated to breathing activity is laminar, no turbulence has been considered in the numerical investigations. The droplet motion has been described using a solving approach based on the following LPT equation:

$$m_d \frac{d\boldsymbol{u}_d}{dt} = \boldsymbol{F}_D + \boldsymbol{F}_g \qquad (1)$$

and

$$\frac{dx_d}{dt} = \boldsymbol{u}_d \qquad (2)$$

where $m_d$ $(kg)$ is the mass of the droplet; $\boldsymbol{u}_d\left(\frac{m}{s}\right)$ represents the droplet velocity; $t$ $(s)$ is the time; $\boldsymbol{F}_D$ $(N)$ and $\boldsymbol{F}_g(N)$ are, respectively, the drag and gravity forces acting on the droplet; and $x_d$ $(m)$ represents the trajectory of the droplet. The drag force is given by Crowe (2011):

$$F_D = m_d \frac{18}{\rho_d \cdot d_d^2} C_D \frac{Re_d(\boldsymbol{u} - \boldsymbol{u}_d)}{24} \qquad (3)$$

In eq. (3), $\rho_d\left(\frac{kg}{m^3}\right)$, $d_d(m)$ and $Re_d$ represent, respectively, the density, diameter and Reynolds number of the droplet. The droplet density has been considered constant and equal to 1200 kg m$^{-3}$. The $Re_d$ is calculated as:

$$Re_d = \frac{\rho(|\boldsymbol{u} - \boldsymbol{u}_d|)d_d}{\mu} \qquad (4)$$

where $\rho\left(\frac{kg}{m^3}\right)$ is the air density, whereas the drag coefficient, $C_D$, in equation (3) is evaluated as a function of the droplet Reynolds number:

$$C_D = \begin{cases} \frac{24}{Re_d} & \text{if } Re_d < 1 \\ \frac{24}{Re_d}(1 + 0.15 \cdot Re_d^{0.687}) & \text{if } 1 \leq Re_d \leq 1000 \\ 0.44 & \text{if } Re_d > 1000 \end{cases} \qquad (5)$$

Droplet collisions are considered to be elastic, and the equations of motion for the droplets are solved assuming a two-way coupling: the flow field affects the droplet motion and vice-versa.

*2.2. Scenario analyzed: close contact during speaking*

The Eulerian-Lagrangian based model described in section 2.1 has been applied to the analysis of droplet dispersion in close contact during speaking. In particular, face-to-face interactions between two subjects (infected emitter and susceptible receiver) of the same height, located at different distances in the range 0.25–1.75 m, were studied. The susceptible subject was considered to be a mouth-breather, thus airborne droplets were inhaled through the mouth. Moreover, because oral, nasal (Gallo et al., 2021), and ocular mucosa (Lu et al., 2020) have been recognized as possible transmission routes for respiratory viruses, large droplet deposition onto mouth, ocular and nostril surfaces due to their inertial trajectories was estimated through the Eulerian-Lagrangian numerical model.

In Figure 1, the computational domain including the external surfaces, the emitter, and the receiver is illustrated. The cad file for emitter and receiver was obtained as an opensource file by the website "gradcad.com". The mathematical model described in section 2.1 was numerically solved using the open source openFOAM software, based on the finite volume technique, under the boundary conditions presented in Table 1.

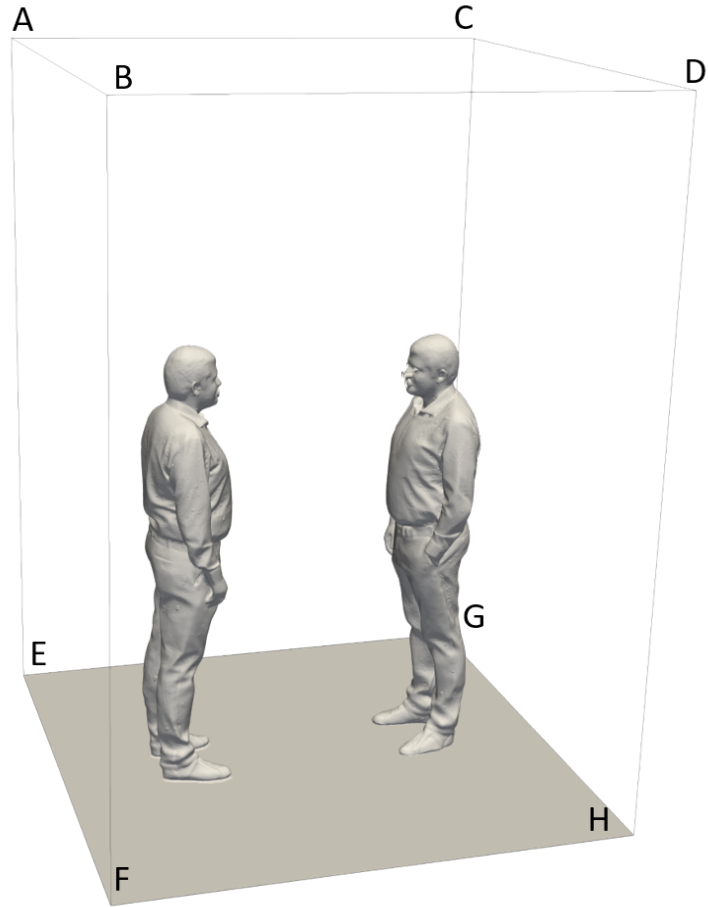

**Figure 1** - Computational domain in which the emitter (on the left), the receiver (on the right), and the external surfaces have been highlighted.

**Table 1** - Boundary conditions (BC) adopted for the external surfaces of the computational domain and for the subjects.

| Surface | BC for velocity | BC for pressure | BC for temperature |
| --- | --- | --- | --- |
| ABEF | $\boldsymbol{u} = 0$ | $\frac{\partial \boldsymbol{p}}{\partial n} = 0$ | $\frac{\partial T}{\partial n} = 0$ |
| CDGH | $\boldsymbol{u} = 0$ | $\frac{\partial \boldsymbol{p}}{\partial n} = 0$ | $\frac{\partial T}{\partial n} = 0$ |
| BDFH | $\boldsymbol{u} = 0$ | $\frac{\partial \boldsymbol{p}}{\partial n} = 0$ | $\frac{\partial T}{\partial n} = 0$ |
| ACEG | $\boldsymbol{u} = 0$ | $\frac{\partial \boldsymbol{p}}{\partial n} = 0$ | $\frac{\partial T}{\partial n} = 0$ |
| ABCD | $\frac{\partial \boldsymbol{u}}{\partial n} = 0$ | $\boldsymbol{p} = 101325\ Pa$ | $T = 293.15\ K$ |
| Emitter mouth | see Figure 2 | $\frac{\partial \boldsymbol{p}}{\partial n} = 0$ | $T = 308.15\ K$ |
| Receiver mouth | $\boldsymbol{u} = A \cdot sin(2\pi f t)$ | $\frac{\partial \boldsymbol{p}}{\partial n} = 0$ | $T = 308.15\ K$ |

In Figure 2, the schematization of the surfaces of interest of the two subjects is presented: the emitter and receiver mouths were modeled as circular surfaces with a diameter of 2 cm, the nostrils were modeled as circular surfaces with a diameter of 1.13 cm, and the eyes were modeled as ellipses with axes of 2.76 and 1.38 cm, respectively (Chao et al., 2009; Chen et al., 2020).

In Figure 2, the boundary condition in terms of air velocity at the mouths of the emitter and receiver is also graphed; in particular, a sinusoidal approximation of breathing is adopted to realistically simulate a real interaction between two subjects. The volumetric flow rates were selected as the average values among those indicated by Abkarian et al. (2020): 1 L s$^{-1}$ for speaking and 0.45 L s$^{-1}$ for mouth breathing. In particular, the transient sinusoidal velocity profile applied at the receiver mouth presents an amplitude of 1 m s$^{-1}$ and a frequency of 0.2 s$^{-1}$, assuming a time period of 5 s for a full breath. The amplitude value was selected on the basis of the PIV measurement results reported in section 3.1. Velocity peaks of 5 m s$^{-1}$ mounted on the sinusoidal velocity profile were considered for the emitter during speaking, as confirmed by Abkarian et al. (2020) and by the PIV experimental analysis for speaking (see PIV results in section 3.1). As concerns the velocity vector direction from the emitter's mouth, a conical jet flow was considered, adopting a cone angle equal to 22° with random velocity directions in intervals of 0.1 s. This adopted angle was calculated by Abkarian et al. (2020) to enclose 90% of the particles in a cone passing through the mouth exit, and was verified to remain stable with time after the initial cycles.

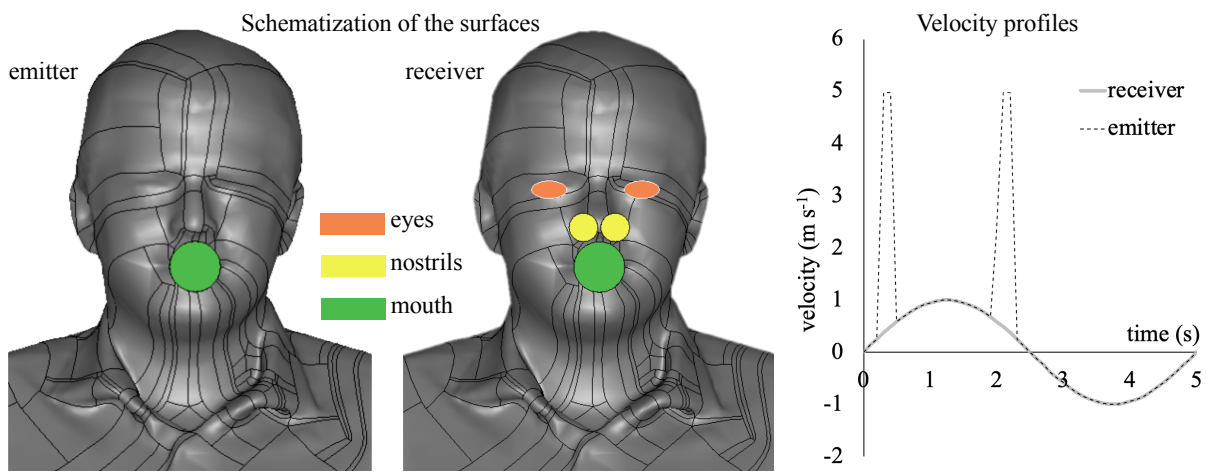

**Figure 2** - Schematization of the surfaces of interest for emitter and receiver (eyes, nostrils, and mouth) and the transient velocity profile adopted as a boundary condition at the emitter and receiver mouths.

Careful attention was paid to the computational mesh construction; in particular, simulations were performed employing hexahedral-based unstructured computational grids, and realized employing the open source snappyHexMesh algorithm, chosen on the basis of a proper mesh sensitivity analysis. In particular, three meshes were selected: Mesh 1 composed of 687380 elements, Mesh 2 composed of 1801060 elements, and Mesh 3 composed of 3023827 elements. The average percent deviation amongst the velocity fields obtained comparing Mesh 1 and Mesh 2 was equal to 6.56%, while comparing Mesh 2 and Mesh 3 resulted in an average percent deviation equal to 1.93%. Because the percent deviation amongst the velocity fields obtained between Mesh 2 and Mesh 3 was low (lower than 5%), the simulations were carried out adopting Mesh 2. As an example, Figure 3 shows the computational grid employed (Mesh 2) to simulate droplet spread in the case of an interpersonal distance of 0.76 m. The grid is refined in correspondence of the solid surfaces, where a boundary layer region is added to better capture the viscous region gradients, and presents a maximum non-orthogonality value of about 50.

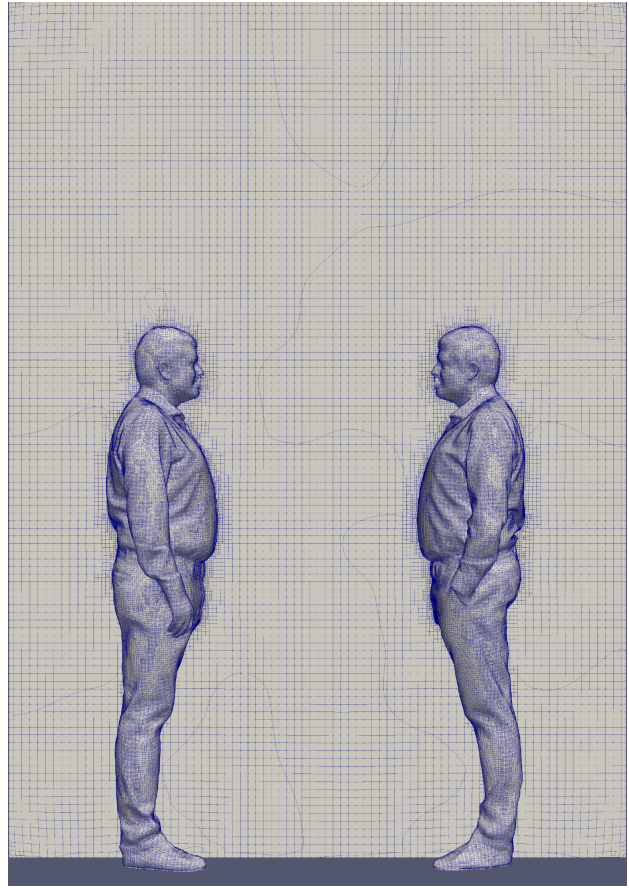

**Figure 3 -** Computational grid employed (Mesh 2, 1801060 elements) to simulate droplet spread in the case of an interpersonal distance of 0.76 m**.**

*2.3. Particle image velocimetry experimental investigations*

PIV measurements, aimed at validating the fluid-dynamic simulation results, were performed at TU Delft's laboratories to study the air flow exhaled during human expiratory activities. The experimental setup (Figure 4) consists of an sCMOS camera from LaVision (2560 × 2160 px) coupled with a Nikon objective lens (35 mm focal length), an Nd:YAG Quantel laser (Evergreen, 200 mJ per pulse) and a smoke generator. The laser sheet (2–3 mm thickness) was formed from below the mouth, and passed through the subject's mid-plane, whereas the sCMOS camera was positioned approximately at the subject's mouth height, at a distance of 80 cm from the laser sheet, with the objective's axis perpendicular to it. The image magnification was 0.05, rendering a field of view of 30 cm (height) × 36 cm (width), while the resolution was 0.14 mm px$^{-1}$. Images were acquired in frame-straddling mode (double frame, single exposure) at 10 Hz, with a time interval of 500 µs between frames.

The subject (male, 32 years old, 1.84 m, 80 kg) was protected against the laser light by safety goggles and a black screen positioned in front of him (not shown), with a 5 cm diameter opening for the mouth-exhaled air. A 3 cm long cylinder of the same diameter was placed at the opening to help position the head and to block the laser light from below. The head was positioned with the subject's nose slightly touching the upper surface of the cylinder; therefore, inhalation and exhalation through the nose did not influence the measured flow velocities. The entire setup, including the subject, was encompassed by a black tent (about 15 m$^3$), whose main objective was to contain the smoke. The entire tent was filled with smoke by turning the smoke generator on for about 2 s with the tent closed, and waiting for about 10 min for the smoke to become homogeneously spread and the flow disturbances due to the smoke generator to become negligible.

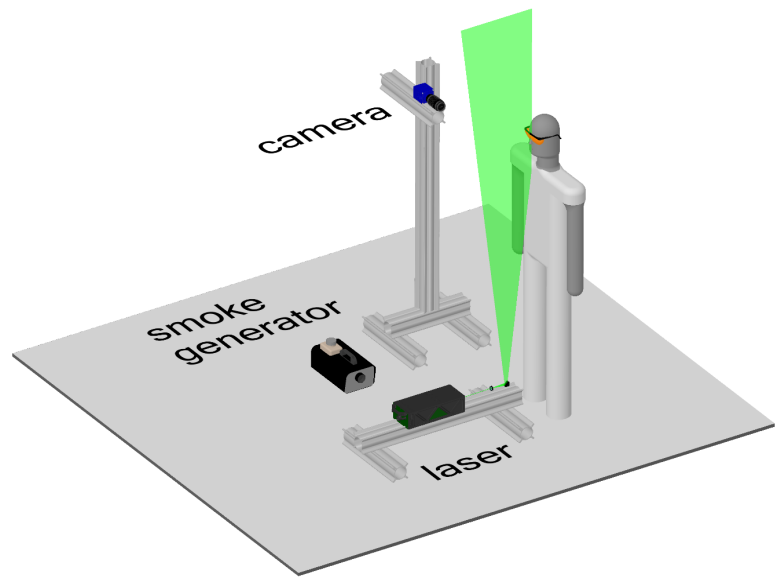

**Figure 4** - Particle image velocimetry experimental setup.

Three different respiratory activities were investigated: inhaling through the nose and exhaling from the mouth, inhaling through and exhaling from the mouth, and speaking. Each activity was recorded for a duration of 50 s (500 images), which comprised about five respiratory cycles. The speaking activity consisted of reciting an excerpt from the rainbow passage (Fairbanks, 1941), a speech often used for the study of voice and articulation and representative of the multiple sounds of the English language: *"When the sunlight strikes raindrops in the air, they act as a prism and form a rainbow. The rainbow is a division of white light into many beautiful colours. These take the shape of a long round arch, with its path high above, and its two ends apparently beyond the horizon. There is, according to legend, a boiling pot of gold at one end. People look, but no one ever finds it. When a* *man looks for something beyond his reach, his friends say he is looking for the pot of gold at the end of the rainbow."*

The images were processed via cross-correlation analysis, using the software DaVis 8.4 from LaVision. The final interrogation window was 48 × 48 px (7 × 7 mm$^2$) with 75% overlap, yielding about 160 × 200 vectors per image. Typical uncertainty of a PIV displacement measurement is 0.1 px (Raffel et al., 2018); because the velocity magnitude close to the mouth varied in the range of 1–5 m s$^{-1}$ (3–18 px), the uncertainty of the instantaneous velocity is estimated to be within 0.5%–3%.

*2.4. Droplet emission*

The number of droplets exhaled by the infected subject as a function of the diameter per unit time, i.e. the droplet number emission rate (ER$_N$, droplet s$^{-1}$), was estimated starting from the number distribution of the droplets emitted by an adult person provided by (Johnson et al., 2011; Morawska et al., 2009). They measured the droplet distribution from 0.5 μm to about 1000 μm in close proximity of an adult person's mouth while speaking, in order to consider negligible the droplet evaporation phenomenon. Such measurement was extremely challenging; indeed, the experimental analysis was performed in a purpose build wind tunnel (named the expired droplet investigation system, EDIS) applying two separate measurement techniques to cover the entire size range: an aerodynamic particle sizer (up to 20 μm) and a droplet deposition analysis (20–1000 μm). For the sake of brevity, the experimental analyses performed in that study are not exhaustively described here; interested readers can refer to the original papers for further details.

To make the simulations affordable, the droplet distributions (Johnson et al., 2011; Morawska et al., 2009) were fitted through a simplified distribution. In particular, from the number distribution provided by (Johnson et al., 2011; Morawska et al., 2009), the volume distribution was calculated considering spherical droplets, then both number and volume distributions were fitted through simplified distributions made up of seven diameters (i.e. seven size ranges). Because the evaporation phenomenon occurs quickly as soon as the droplets are emitted (Balachandar et al., 2020; Xie et al., 2007), in the present paper the post-evaporation number and volume distributions were considered in the simulation. To this end, the volume droplet distribution before evaporation (i.e. as emitted) was reduced to that resulting from the quick evaporation, which is the volume fraction of non-volatiles in the initial droplet, here considered equal to 1% (Balachandar et al., 2020). Therefore, the droplet shrinkage due to evaporation reduces the droplet diameter to about 20% of the initial emitted size. Additionally, the shrinking effect is not homogeneous for the entire size range. In particular, as reported in the scientific literature (Balachandar et al., 2020; Xie et al., 2007), the evaporation is slow for very small droplets (< 1 μm) and quite negligible for large droplets. Therefore, we (i) grouped all the droplets < 1 μm after the evaporation (i.e. droplets < 4.6 μm at emission) in a single size interval labelled as 1 μm droplet diameter; (ii) considered the droplet nuclei resulting from the evaporation process for droplets with an initial diameter of 4.6–90 μm (reduced to droplets with a diameter of 1–19.2 μm after evaporation); and (iii) neglected the evaporation for droplets > 90 μm at emission. The resulting number and volume distributions are summarized in Table 2 and in Figure 5. In Table 2 the resulting droplet number ($ER_N$) and volume ($ER_V$, pre-evaporation) emission rates are also reported, calculated by multiplying the number (or volume) concentration at each size by the expiration flow rate of a speaking subject while standing (1.0 L s$^{-1}$, average value measured for an adult by Abkarian et al. [2020]). The total droplet number and volume concentrations are reported in Table 2: the total droplet number concentration (0.25 droplet cm$^{-3}$) is the same before and after evaporation, whereas a small variation was recognizable for the volume concentration (6.27 × 10$^{-5}$ and 6.19 × 10$^{-5}$ μL cm$^{-3}$ before and after evaporation, respectively) due to shrinkage of droplets initially < 90 μm. In terms of the number concentration (or emission rate), the contribution of the airborne droplets is 98%, whereas it is only 1% and 0.01% in terms of volume concentration (or emission rate) before and after evaporation, respectively, thus confirming that a reduced number of large droplets mostly contributes to the total volume emitted.

**Table 2** - Droplet number and volume distributions pre- and post-evaporation (fitted by seven size ranges) adopted in the simulations. Droplet diameters and corresponding ranges before and after evaporation are also reported, as well as droplet number and volume emission rates. Airborne and large droplets were separately identified.

| Type of droplets | Pre-evaporation | | | | | Post-evaporation | | |
|---|---|---|---|---|---|---|---|---|
| | Droplet diameter, $d_d$ (μm) | dN/dlog($d_d$) (droplet cm$^{-3}$) | dV/dlog($d_d$) (μL cm$^{-3}$) | $ER_N$ (droplet s$^{-1}$) | $ER_V$ (μL s$^{-1}$) | Droplet diameter, $d_d$ (μm) | dN/dlog($d_d$) (droplet cm$^{-3}$) | dV/dlog($d_d$) (μL cm$^{-3}$) |
| Airborne droplets | 4.6 μm (< 0.5 to 4.6 μm) | 0.266 | 1.39×10$^{-10}$ | 217.6 | 1.14×10$^{-7}$ | 1 μm (< 1 μm) | 0.266 | 1.39×10$^{-8}$ |
| | 9.0 μm (4.6 to 17.7 μm) | 0.035 | 1.33×10$^{-10}$ | 20.3 | 7.80×10$^{-8}$ | 1.9 μm (1.0 to 3.8 μm) | 0.035 | 1.33×10$^{-8}$ |
| | 23.2 μm (17.7 to 30.4 μm) | 0.013 | 8.74×10$^{-10}$ | 3.1 | 2.05×10$^{-7}$ | 5 μm (3.8 to 6.6 μm) | 0.013 | 8.74×10$^{-8}$ |
| | 45.5 μm (30.4 to 68.2 μm) | 0.016 | 8.08×10$^{-9}$ | 5.7 | 2.83×10$^{-6}$ | 9.8 μm (6.6 to 14.7 μm) | 0.016 | 8.08×10$^{-7}$ |
| | 78 μm (68 to 90 μm) | 0.015 | 3.83×10$^{-8}$ | 1.8 | 4.48×10$^{-6}$ | 16.8 μm (14.7 to 19.2 μm) | 0.015 | 3.83×10$^{-6}$ |
| Large droplets | 100 μm (90 to 120 μm) | 0.014 | 7.09×10$^{-6}$ | 1.6 | 8.29×10$^{-4}$ | 100 μm (90 to 120 μm) | 0.014 | 7.09×10$^{-6}$ |
| | 300 μm (120 to 800 μm) | 0.005 | 6.53×10$^{-5}$ | 4.3 | 6.11×10$^{-2}$ | 300 μm (120 to 800 μm) | 0.005 | 6.53×10$^{-5}$ |
| Total | | 0.254 | 6.27×10$^{-5}$ | 254.5 | 6.19×10$^{-2}$ | | 0.254 | 6.19×10$^{-5}$ |

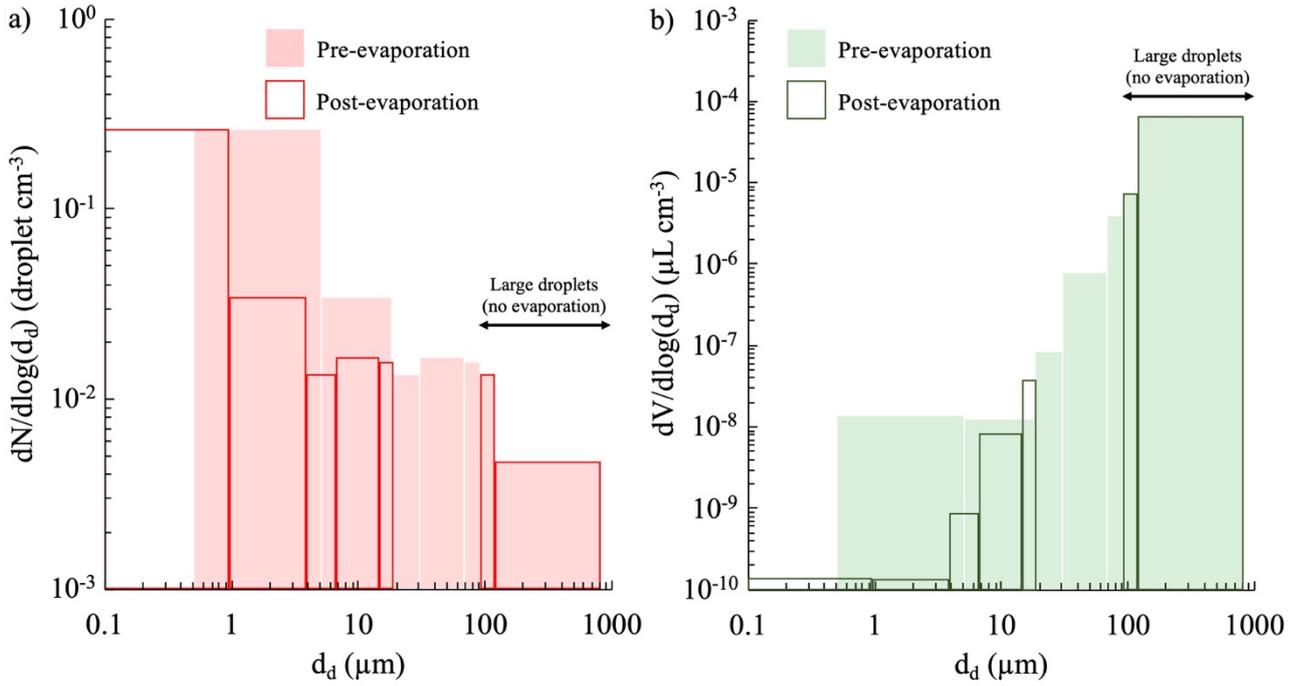

**Figure 5** - Droplet number (a) and volume (b) distributions adopted in the simulations as fitted through seven size ranges; in particular, distributions pre- and post-evaporation are reported to show how airborne and large droplets are affected by the evaporation phenomenon.

*2.5. Estimation of the dose received by the susceptible subject and infectious risk assessment*

The viral load carried by the droplets exhaled by the infected subject was evaluated as the product of the droplet volume (discussed in the previous section) and the corresponding viral load. The viral load of an infected subject, $c_v$, can vary significantly (several orders of magnitude) (Buonanno et al., 2020a, 2020b; Mikszewski et al., 2021); thus, to achieve a proper infection risk assessment of an exposed subject, all the possible viral load data should be considered. In other words, when calculating the dose of RNA copies received by the susceptible subject (through inhalation or deposition), the probability distribution function of $c_v$ values should be considered, which is the probability of occurrence of each $c_v$ value. Data of the viral load in sputum so far available in the scientific literature (Fajnzylber et al., 2020; Pan et al., 2020; Wölfel et al., 2020) can be fitted through a log-normal distribution characterized by average and standard deviation $c_v$ values of $\log_{10} 5.6$ and $\log_{10} 1.2$ RNA copies mL$^{-1}$ (Mikszewski et al., 2021), i.e. 1$^{st}$, 50$^{th}$, and 99$^{th}$ percentiles equal to $6.4 \times 10^2$, $4.0 \times 10^5$, and $2.5 \times 10^8$ RNA copies mL$^{-1}$, respectively.

The large and airborne droplet doses of RNA copies ($D_{large}$ and $D_{airborne}$) received by the susceptible subject for each $c_v$ value were calculated as:

$$D_{large}(c_v) = \int_0^T \left(V_{d-large}(t) \cdot c_v\right) dt \qquad (6)$$

$$D_{airborne}(c_v) = \int_0^T \left(V_{d-airborne-pre}(t) \cdot c_v\right) dt$$

where $V_{d\text{-}large}(t)$ and $V_{d\text{-}airborne\text{-}pre}(t)$ are the doses of airborne droplets inhaled and large droplets deposited as a function of the exposure time ($t$), and $T$ is the total exposure time. The authors point out that the viral load carried by the droplet is related to the initial droplet volume (i.e. before evaporation), and evaporation leads to a reduction in the droplet volume. The RNA copies do not evaporate; thus, the $V_{d\text{-}airborne\text{-}pre}$ term refers to the dose of airborne droplets calculated with the initial (pre-evaporation) volume. The $V_{d\text{-}airborne\text{-}pre}$ term has been adopted to distinguish it from the actual doses of airborne droplets inhaled ($V_{d\text{-}airborne\text{-}post}$), i.e. droplets with the actual volume at the time of inhalation (i.e. post-evaporation). The total dose of RNA copies received by the exposed subject for each $c_v$ value was then evaluated by summing up the deposition and inhalation contributions, i.e. $D_{total}(c_v) = D_{large}(c_v) + D_{airborne}(c_v)$.

From the dose of RNA copies, the probability of infection ($P_I$) of the exposed subject for each $c_v$ was calculated adopting a well-known exponential dose–response model (Haas, 1983; Sze To and Chao, 2010; Watanabe et al., 2010):

$$P_I(c_v) = 1 - e^{-\frac{D_{total}(c_v)}{HID_{63}}} \qquad (\%) \qquad (7)$$

where $HID_{63}$ represents the human infectious dose for 63% of susceptible subjects, i.e. the number of RNA copies needed to initiate the infection with a probability of 63%. For SARS-CoV-2, a $HID_{63}$ value of $7 \times 10^2$ RNA copies was adopted based on the thermodynamic-equilibrium dose–response model developed by Gale (2020).

The individual risk of infection ($R$) of the exposed person was then calculated by integrating, for all the possible $c_v$ values, the product between the conditional probability of the infection for each $c_v$ ($P_I(c_v)$) and the probability of occurrence of each $c_v$ value ($P_{cv}$):

$$R = \int_{c_v} (P_I(c_v) \cdot P_{cv}) dc_v \qquad (\%) \qquad (8)$$

The authors point out that in the present analysis an equal amount of RNA copies received by inhalation of airborne droplets or by deposition of large droplets was considered to cause the same effect in terms of infection.

## 3. Results and discussion

*3.1. Particle image velocimetry measurements and numerical results*

As mentioned in the methodology section, PIV measurement results for a mouth breathing case study provided the required information to choose the velocity boundary conditions employed in the computational fluid dynamics (CFD) numerical simulations summarized in section 2.2.

In particular, the adopted boundary condition for the mouth-breathing receiver was verified by comparing numerical results with PIV data in terms of velocity profiles obtained at different distances from the mouth of the emitter. This comparison also provided a rough confirmation of the numerical velocity field. To this end, experimental (PIV) and numerical (CFD) velocity contours obtained in the sagittal plane, by synchronizing the instant of time for breathing at which the maximum velocity values are reached, are presented in Figure 6, whereas PIV and CFD vertical velocity profiles in sagittal plane at a distance from the emitter mouth equal to 0.10 m and 0.32 m are compared in Figure 7. The peak numerical and experimental peak velocities differ by 6% and 7% at interpersonal distances of 0.10 m and 0.32 m, respectively, thus validating the numerical solutions obtained through the CFD analyses.

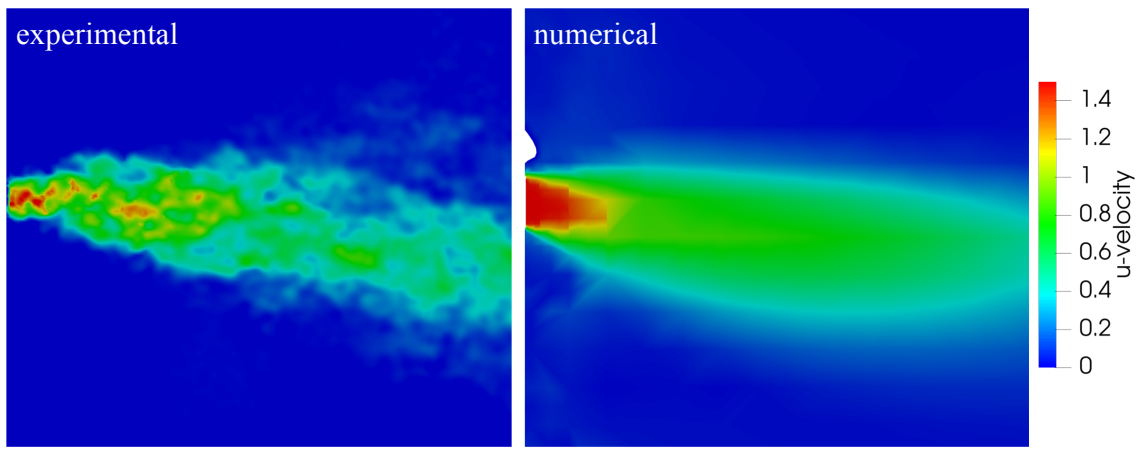

**Figure 6** - Experimental and CFD velocity contours obtained in a sagittal plane by synchronizing the instant of time for breathing at which the maximum velocity values are reached.

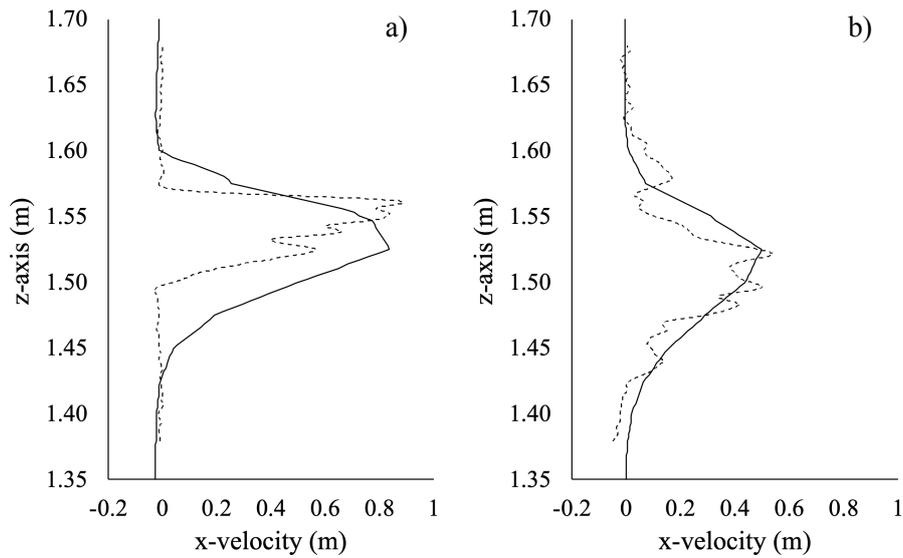

**Figure 7** - Experimental (particle image velocimetry, dotted lines) and CFD (solid lines) velocity profile comparison obtained in a sagittal plane at a distance from the emitter mouth equal to 0.10 m (a) and 0.32 m (b).

The emitter velocity peaks of 5 m s$^{-1}$ adopted in the simulations were confirmed by the experimental analysis related to the speaking expiratory activity. Among the 500 recorded images obtained in the 50 s of the experiment (see section 2.3), the time instant giving the maximum *u*-velocity and *v*-velocity values were selected and illustrated: *u*-velocity peaks of 5 m s$^{-1}$ are clearly recognizable in Figure 8a.

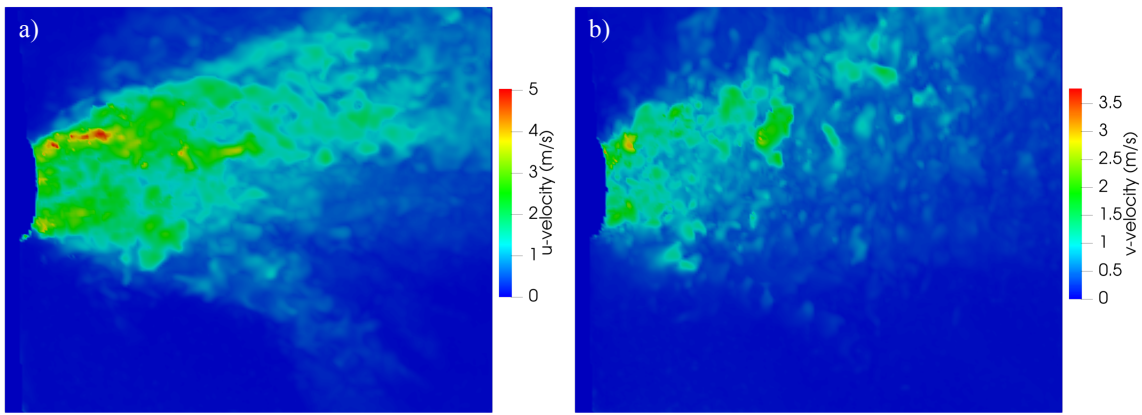

**Figure 8** - Instantaneous *u*-velocity (a) and *v*-velocity (b) contours obtained by particle image velocimetry during reading of the excerpt from the rainbow passage.

*3.2. Droplet dose received by the susceptible subject*

As an illustrative example of the droplet trajectories and flow fields obtained from the simulations, Figure 9 shows the velocity contours and droplet positions for a 5-s breathing period (at computational times of 5, 5.5, 6.5, 7.5, 8.5, and 10 s) in the case of an interpersonal distance of 0.76 m between the injector and receiver mouths. For this distance, large droplets fall to the ground without reaching the susceptible surfaces of the receiver, while the airborne droplets are partly inhaled by the receiver. Indeed, airborne droplets are transported by the air velocity field, reach the receiver, and then are spread while rising in a vertical direction due to the effect of buoyancy forces. In fact, from the analysis of the three dimensional transient air velocity field shown in Figure 9, it can be observed that when the air velocity from the emitter is low (Figure 9a, d, e, f), the effect of buoyancy is evident, while forced convection dominates when the air velocity from the emitter is high (Figure 9b, c).

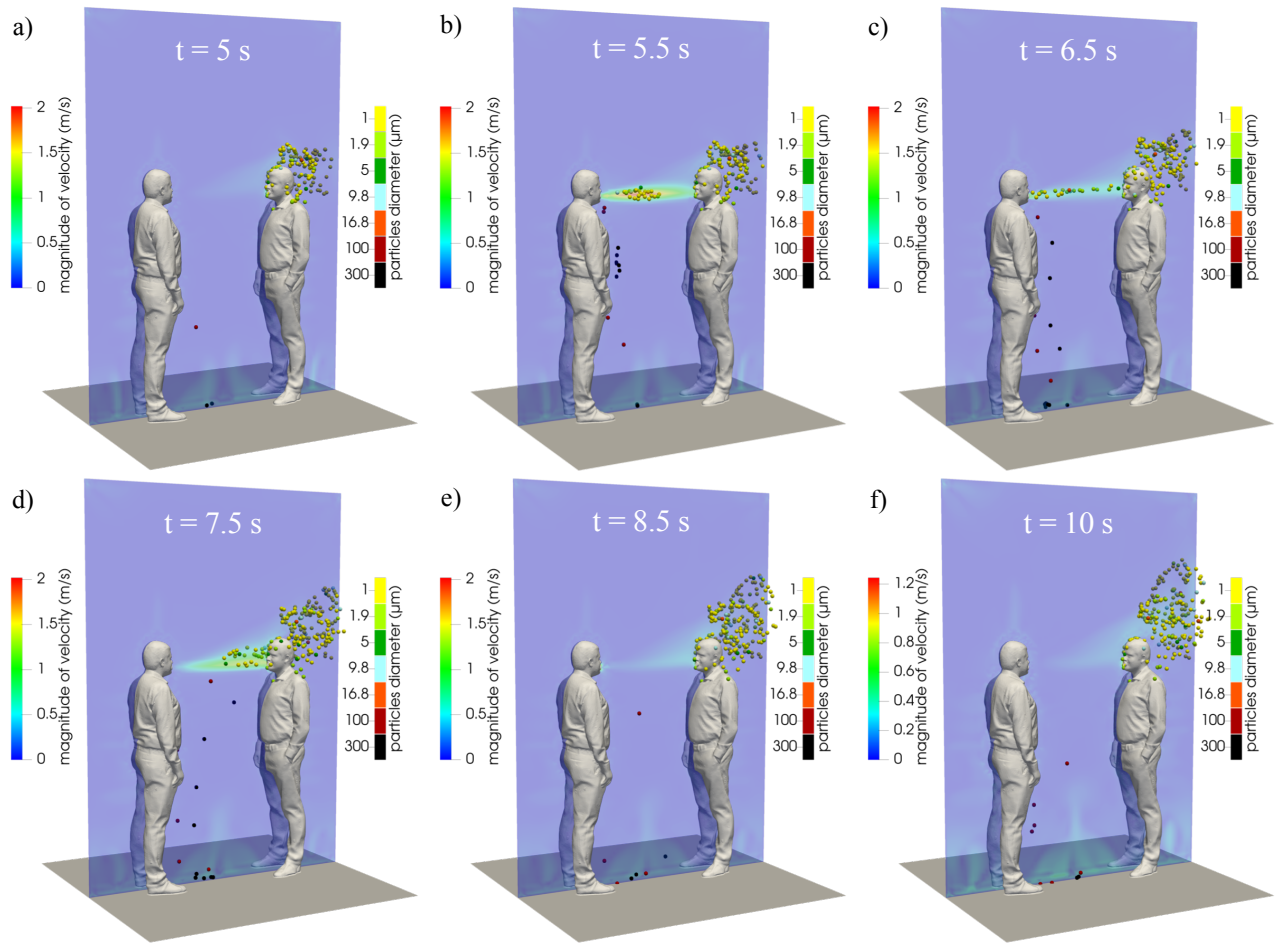

**Figure 9** - Numerical velocity contours during a single breath at a distance of 0.76 m between people: six selected computational times (5, 5.5, 6.5, 7.5, 8.5, and 10 s) are shown.

Figure 10 shows airborne and large droplet doses (i.e. $V_{d\text{-}airborne\text{-}pre}$, $V_{d\text{-}airborne\text{-}post}$, and $V_{d\text{-}large}$, μL) as a function of distance for an exposure time of 1 min. Data were obtained by performing numerical simulations of 15 min and averaging the obtained volumes over an observation time equal to 1 min. The trends show that the large droplet dose dominates up to a distance of < 0.6 m but, beyond this distance, a step-decrease is observed because the large droplets cannot reach the deposition surfaces of the susceptible subject due to their inertial trajectories. Figure 10 also shows the dose of non-evaporated airborne droplets ($V_{d\text{-}airborne\text{-}pre}$); this information is useful because infectivity (and therefore the risk) is directly related to this metric. For distances > 0.6 m, only the airborne droplet contribution to the total dose received by the infected subject is observed. To summarize, the interpersonal distance is a key parameter in evaluating the close contact risk because the susceptible subject could fall within the highly concentrated droplet-laden flow exhaled by the infected subject.

While the trajectory of large droplets is mostly affected by their inertia, and the related effect is negligible for distances > 0.6 m, the spread of airborne droplets is affected by the spread angle of the exhaled flow. In particular, we recognized that at short interpersonal distances (roughly < 0.76 m) from the emission point, where the exhaled air flow angle is still narrow, the dose of airborne droplets decays following the 1/L rule (with L representing the interpersonal distance), whereas for interpersonal distances in the range of 0.76–1.75 m, where the exhaled air flow angle becomes wider, the dose of airborne droplets decays following the $1/L^2$ rule as recognized for passive tracer-gas decay and reported by Li (2021b).

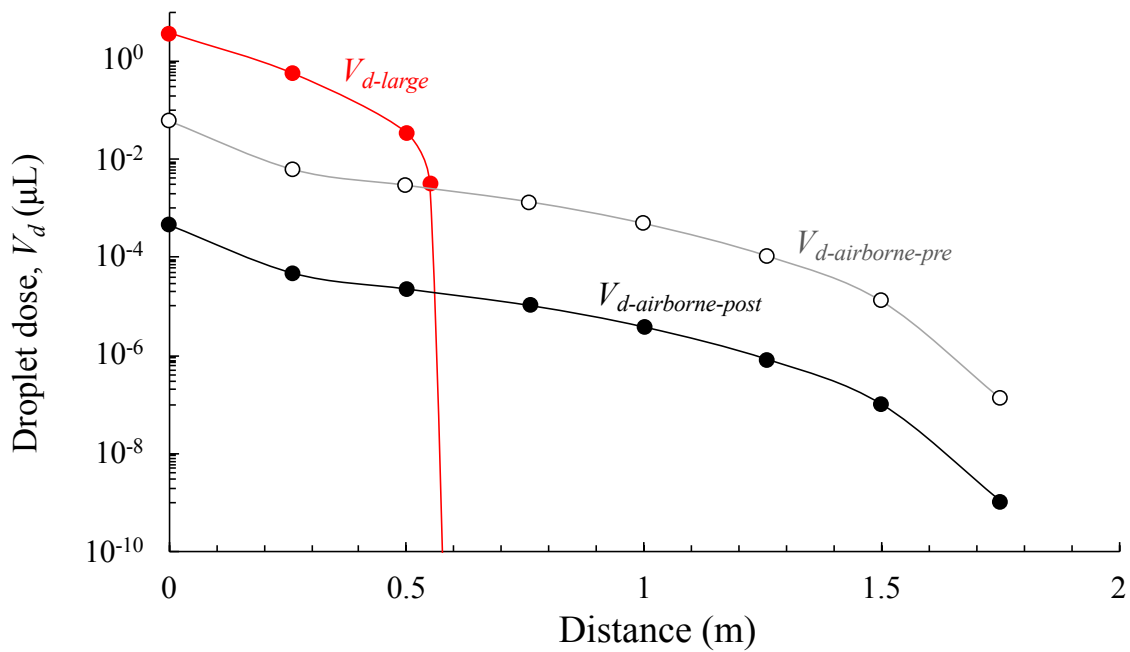

**Figure 10** - 1-min large ($V_{d-large}$) and airborne droplet doses ($V_{d-airborne-pre}$ and $V_{d-airborne-post}$) received by the susceptible subject (by deposition and inhalation, respectively) as a function of the distance between the two subjects.

*3.3. Risk assessment at close contact*

Figure 11 shows the infection risk (R) as a function of the interpersonal distance between the speaking infected subject and the susceptible subject for different exposure times (10 s, 1 min, 15 min). The infection risk is close to 100% for 15-min exposures when the distance is less than 0.6 m, and is extremely high (> 30%) even for short exposures (10-s exposure). In fact, as shown in the previous section, for such short distances the dilution is not sufficient to reduce the deposited and inhaled doses due to large droplets and airborne droplets. Beyond 0.6 m, because there is only the contribution of airborne droplets (inhaled dose), a sharp risk decay is observed in particular for short exposures. Thus, once again, the exposure duration and the interpersonal distance can reduce the infectious risk by several orders of magnitude. As an example, Figure 11 shows that the distance to be adopted to achieve a risk lower than 0.1% in the case of an exposure time of 15 min is around 1.5 m, which is reduced to 1 m for an exposure time of 1 min and 0.75 m for a 10-s exposure. Once again, we point out that the estimated infection risk value is related to an outdoor environment with stagnant air or to an indoor environment without the contribution due to the accumulation of viral load in the environment itself. Nonetheless, when people share indoor air maintaining distancing because of airborne droplets, an infection risk can still be present. For example, in Figure 11 the infection risk is estimated when sharing room air and maintaining distancing on the basis of the zero-dimensional well-mixed approach reported in (Buonanno et al., 2020a, 2020b) and (Mikszewski et al., 2021) in small (60 m$^3$, e.g. offices, classrooms) and large volumes (400 m$^3$, e.g. restaurants, conference rooms). The simulations were performed for an exposure time of 15 min for typical ventilation rates (0.2–3.0 h$^{-1}$) occurring in indoor environments (Frattolillo et al., 2021; Stabile et al., 2019). The 15-min close contact (spatially-dependent) infection risk merges into the constant (not spatially dependent) sharing room air infection risks at interpersonal distances of about 1.4–1.6 m (depending on the volume and the ventilation rate) as also indicated in Li (2021b); this distance is representative of the boundary of simplified well-mixed model applicability. Thus, for the investigated scenario, infection risk assessments through complex three-dimensional and transient CFD models are essential for interpersonal short distances (< 1.4–1.6 m), whereas simplified zero-dimensional well-mixed models can be applied for longer distances.

When applying these findings, an obvious question arises regarding the typical exposure duration and distance data. To address this issue, Zhang et al. (2020b) monitored and analyzed indoor human behavior in a graduate student office using automatic devices. They measured a median duration of close contact of 15 s and an average interpersonal distance of 0.81 m during such close contacts. Adopting such median exposure durations, the corresponding infection risk is negligible (i.e. less than 0.1%, adopted as the threshold value by Buonanno et al. (2020a), reaching 0.3% for exposure times of 1 min. Only with long exposure times (15 min) would the risk become significantly higher than 1%. Therefore, even though Zhang et al. (2020b) verified that 9.7% of employees' time in offices was in close contact (with 4.0 close contacts h$^{-1}$), an interpersonal distance of 0.81 cm is sufficient to have a limited risk for the measured exposure times (< 1 min) in the analyzed common occupational scenario. Apart from workplace scenarios, the interpersonal distance is an essential feature of individuals' social behavior more broadly in relation to their physical environment and social interactions (Hall, 1966). On the basis of the classical proxemic theory (Hall, 1966), interpersonal distances are classified as (i) *public distance* (> 2 m), (ii) *social distance*, during more formal interactions, (iii) *personal distance*, during interactions with friends, and (iv) *intimate distance*, maintained in close relationships, with Southern European, Latin American, and Arabian countries being the so-called "contact cultures", and North America, Northern Europe, and Asian populations being the "noncontact cultures" (Hall, 1966). Sorokowska et al. (2017) reported significant variability in preferred interpersonal distances across countries as a function of certain characteristics of interacting individuals (such as gender or age), cultural differences, and environmental and sociopsychological factors. They reported worldwide interpersonal distance distributions of $0.56 \pm 0.13$ m, $0.81 \pm 0.12$ m, and $1.06 \pm 0.14$ m for intimate, personal and social distance, respectively, which are included in Figure 11 for discussion. In the case of intimate distance, the average infection risk (i.e. the risk corresponding to the average distance value, i.e. 0.56 m) is extremely high, starting from exposure times of 1 min, but even in the case of short exposures (10 s), it is not negligible (> 1%). For personal distances, the average infection risk is negligible for short exposures (< 0.1%), limited for 1-min exposures (< 1%), and high for 15-min exposures. Finally, in the case of social distances, the average infection risk becomes significant only for 15-min exposures.

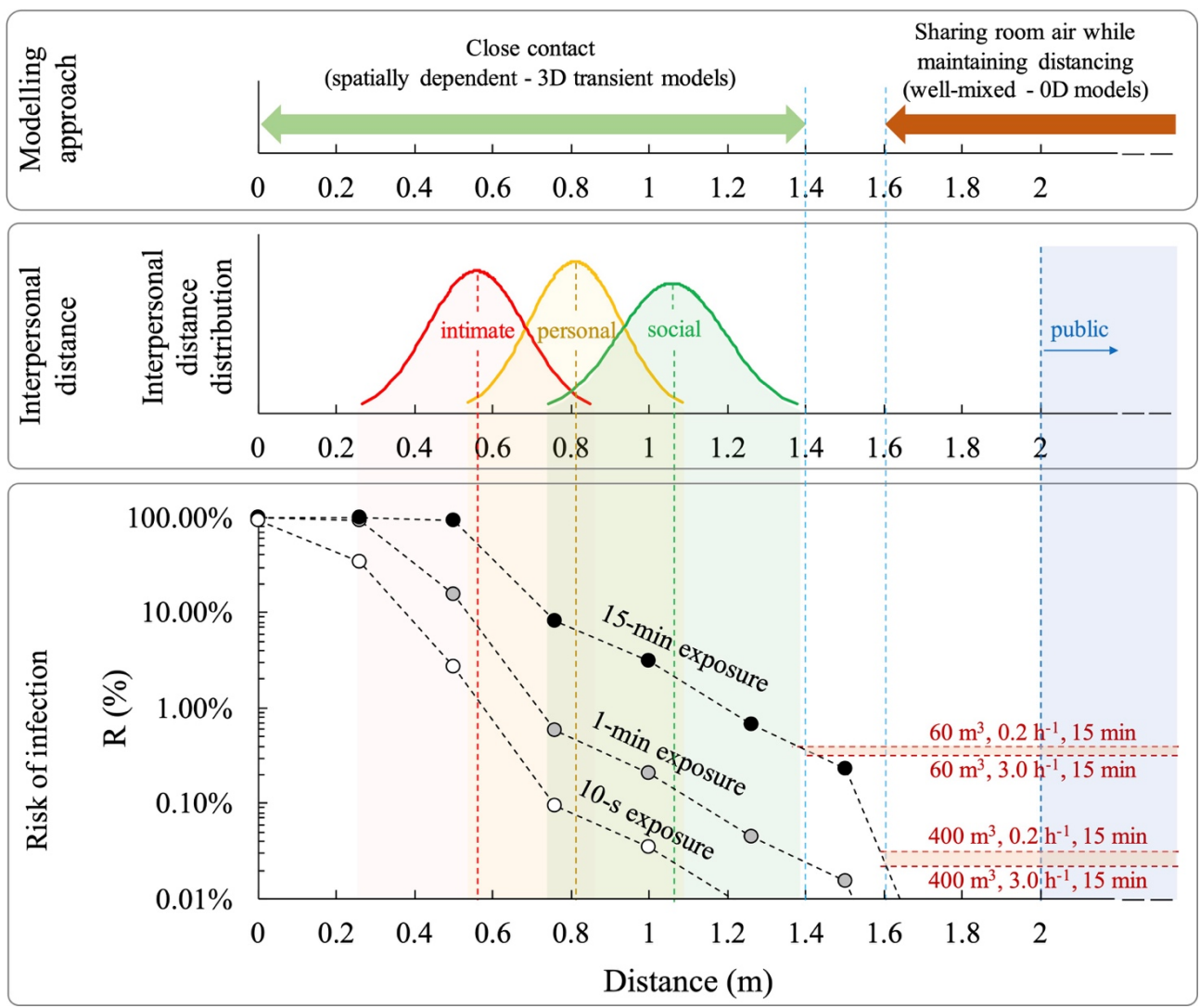

**Figure 11** – Infection risk (R, %) of a susceptible subject as a function of the time of exposure and interpersonal distance from the infected subject; infection risk trends at short and long distances are highlighted as well as the modeling approaches to be applied.

The authors highlight that the approach and the results presented here provide an important insight into potential virus transmission over short distances that could help regulatory authorities and air quality experts in implementing and imposing proper mitigation solutions as a function of the microenvironment and the type of contact expected. The paper also provides information about choosing the proper modeling approach for infection risk assessments. Despite the value of these findings, the authors acknowledge the simplified hypothesis and limitations of this study that should be addressed in future development of this research. First, the results obtained through the CFD approach were compared to the experimental data only in terms of the velocity field and only one subject; second, the effect of turbulence should be investigated for velocity peaks associated to speaking activity; third, the infection was referred to the received droplet dose considering a homogeneous viral load over the entire droplet size range; fourth, the study was limited to mouth-breather subjects; and finally, transmission through surfaces (fomites) was not considered.

## 4. Conclusions

To the authors' knowledge, this is the first study in which an integrated risk assessment is developed for SARS-CoV-2 combining thermo-fluid dynamic modeling and infectious risk assessment to investigate close contact exposure scenarios. The integrated approach relies upon: i) an Eulerian-Lagrangian numerical model for the description of droplets spread once emitted by a speaking person, ii) PIV measurements for the definition of the boundary conditions and to validate the numerical model; iii) definition of a droplet emission model; and iv) infectious SARS-CoV-2 risk assessment. The approach presented here was applied to a close contact between a speaking infected subject and a susceptible subject in a face-to-face orientation with stagnant air conditions.

The results show that the contribution of large droplets (> 100 μm) to the dose and risk received by the susceptible subject is dominant for an interpersonal distance < 0.6 m, which means highly influential on the infection risk only at intimate interpersonal distances (average distance of 0.56 m). In fact, in the case of personal (0.81 m) and social (1.06 m) distancing, the only contribution to the risk of infection comes from airborne droplets. In addition to the distance, the exposure time plays a key role in the risk of infection; in fact, the average infection risk is not negligible even for short exposures (10 s) in the case of intimate distance, whereas in the case of social distances only long exposures (15 min) can lead to a not negligible risk.

A possible threshold value to be adopted as a safe distance in close contact is around 1.5 m, which lowers the infection risk to 0.1% order of magnitude even with prolonged exposure times (15 min). Such a threshold value also represents the boundary distance beyond which simplified well-mixed approaches can be adopted instead of complex spatially dependent three-dimensional transient CFD analyses. Indeed, we have shown that the same infection risk values for short (close contact) and long distances (sharing room air while maintaining distancing) can be obtained for a typical indoor environment for interpersonal distances in the range of 1.4–1.6 m, thus confirming that 1.5 m can be adopted as the typical distance for close contact.

# References

Abkarian, M., Mendez, S., Xue, N., Yang, F., Stone, H.A., 2020. Speech can produce jet-like transport relevant to asymptomatic spreading of virus. Proc. Natl. Acad. Sci. 117, 25237. https://doi.org/10.1073/pnas.2012156117

Ai, Z., Hashimoto, K., Melikov, A.K., 2019. Influence of pulmonary ventilation rate and breathing cycle period on the risk of cross-infection. Indoor Air 29, 993–1004. https://doi.org/10.1111/ina.12589

Ai, Z.T., Melikov, A.K., 2018. Airborne spread of expiratory droplet nuclei between the occupants of indoor environments: A review. Indoor Air 28, 500–524. https://doi.org/10.1111/ina.12465

Arpino, F., Cortellessa, G., Dell'Isola, M., Massarotti, N., Mauro, A., 2014. High order explicit solutions for the transient natural convection of incompressible fluids in tall cavities. Numer. Heat Transf. Part Appl. 66, 839–862. https://doi.org/10.1080/10407782.2014.892389

Balachandar, S., Zaleski, S., Soldati, A., Ahmadi, G., Bourouiba, L., 2020. Host-to-host airborne transmission as a multiphase flow problem for science-based social distance guidelines. Int. J. Multiph. Flow 132, 103439. https://doi.org/10.1016/j.ijmultiphaseflow.2020.103439

Buonanno, G., Morawska, L., Stabile, L., 2020a. Quantitative assessment of the risk of airborne transmission of SARS-CoV-2 infection: Prospective and retrospective applications. Environ. Int. 145, 106112. https://doi.org/10.1016/j.envint.2020.106112

Buonanno, G., Stabile, L., Morawska, L., 2020b. Estimation of airborne viral emission: Quanta emission rate of SARS-CoV-2 for infection risk assessment. Environ. Int. 141, 105794. https://doi.org/10.1016/j.envint.2020.105794

Chao, C.Y.H., Wan, M.P., Morawska, L., Johnson, G.R., Ristovski, Z.D., Hargreaves, M., Mengersen, K., Corbett, S., Li, Y., Xie, X., Katoshevski, D., 2009. Characterization of expiration air jets and droplet size distributions immediately at the mouth opening. J. Aerosol Sci. 40, 122–133. https://doi.org/10.1016/j.jaerosci.2008.10.003

Chapin, C.V., 1912. The sources and modes of infection. J. Wiley & sons; [etc., etc.], New York.

Chen, W., Zhang, N., Wei, J., Yen, H.-L., Li, Y., 2020. Short-range airborne route dominates exposure of respiratory infection during close contact. Build. Environ. 176, 106859. https://doi.org/10.1016/j.buildenv.2020.106859

Crowe, T.C., 2011. Multiphase ows with droplets and particles. CRC Press.

Fairbanks, G., 1941. Voice and Articulation Drillbook. The Laryngoscope, Harper and Brothers 51, 1141–1141. https://doi.org/10.1288/00005537-194112000-00007

Fajnzylber, J., Regan, J., Coxen, K., Corry, H., Wong, C., Rosenthal, A., Worrall, D., Giguel, F., Piechocka-Trocha, A., Atyeo, C., Fischinger, S., Chan, A., Flaherty, K.T., Hall, K., Dougan, M., Ryan, E.T., Gillespie, E., Chishti, R., Li, Y., Jilg, N., Hanidziar, D., Baron, R.M., Baden, L., Tsibris, A.M., Armstrong, K.A., Kuritzkes, D.R., Alter, G., Walker, B.D., Yu, X., Li, J.Z., Abayneh, B.A. (Betty), Allen, P., Antille, D., Balazs, A., Bals, J., Barbash, M., Bartsch, Y., Boucau, J., Boyce, S., Braley, J., Branch, K., Broderick, K., Carney, J., Chevalier, J., Choudhary, M.C., Chowdhury, N., Cordwell, T., Daley, G., Davidson, S., Desjardins, M., Donahue, L., Drew, D., Einkauf, K., Elizabeth, S., Elliman, A., Etemad, B., Fallon, J., Fedirko, L., Finn, K., Flannery, J., Forde, P., Garcia-Broncano, P., Gettings, E., Golan, D., Goodman, K., Griffin, A., Grimmel, S., Grinke, K., Hartana, C.A., Healy, M., Heller, H., Henault, D., Holland, G., Jiang, C., Jordan, H., Kaplonek, P., Karlson, E.W., Karpell, M., Kayitesi, C., Lam, E.C., LaValle, V., Lefteri, K., Lian, X., Lichterfeld, M., Lingwood, D., Liu, H., Liu, J., Lopez, K., Lu, Y., Luthern, S., Ly, N.L., MacGowan, M., Magispoc, K., Marchewka, J., Martino, B., McNamara, R., Michell, A., Millstrom, I., Miranda, N., Nambu, C., Nelson, S., Noone, M., Novack, L., O'Callaghan, C., Ommerborn, C., Osborn, M., Pacheco, L.C., Phan, N., Pillai, S., Porto, F.A., Rassadkina, Y., Reissis, A., Ruzicka, F., Seiger, K., Selleck, K., Sessa, L., Sharpe, A., Sharr, C., Shin, S., Singh, N., Slaughenhaupt, S., Sheppard, K.S., Sun, W., Sun, X., Suschana, E. (Lizzie), Talabi, O., Ticheli, H., Weiss, S.T., Wilson, V., Zhu, A., The Massachusetts Consortium for Pathogen Readiness, 2020. SARS-CoV-2 viral load is associated with increased disease severity and mortality. Nat. Commun. 11, 5493. https://doi.org/10.1038/s41467-020-19057-5

Flügge, C., 1897. Ueber Luftinfection. Z. Für Hyg. Infekt. 25, 179–224. https://doi.org/10.1007/BF02220473

Frattolillo, A., Stabile, L., Dell'Isola, M., 2021. Natural ventilation measurements in a multi-room dwelling: Critical aspects and comparability of pressurization and tracer gas decay tests. J. Build. Eng. 42, 102478. https://doi.org/10.1016/j.jobe.2021.102478

Gale, P., 2020. Thermodynamic equilibrium dose-response models for MERS-CoV infection reveal a potential protective role of human lung mucus but not for SARS-CoV-2. Microb. Risk Anal. 16, 100140–100140. https://doi.org/10.1016/j.mran.2020.100140

Gallo, O., Locatello, L.G., Mazzoni, A., Novelli, L., Annunziato, F., 2021. The central role of the nasal microenvironment in the transmission, modulation, and clinical progression of SARS-CoV-2 infection. Mucosal Immunol. 14, 305–316. https://doi.org/10.1038/s41385-020-00359-2

Haas, C.N., 1983. Estimation of risk due to low doses of microorganisms: a comparison of alternative methodologies. Am. J. Epidemiol. 118, 573–582. https://doi.org/10.1093/oxfordjournals.aje.a113662

Hall, E.T., 1966. The hidden dimension. New York : Doubleday, New York.

Johnson, G.R., Morawska, L., Ristovski, Z.D., Hargreaves, M., Mengersen, K., Chao, C.Y.H., Wan, M.P., Li, Y., Xie, X., Katoshevski, D., Corbett, S., 2011. Modality of human expired aerosol size distributions. J. Aerosol Sci. 42, 839–851. https://doi.org/10.1016/j.jaerosci.2011.07.009

Li, Y., 2021a. Basic routes of transmission of respiratory pathogens—A new proposal for transmission categorization based on respiratory spray, inhalation, and touch. Indoor Air 31, 3–6. https://doi.org/10.1111/ina.12786

Li, Y., 2021b. The respiratory infection inhalation route continuum. Indoor Air 31, 279–281. https://doi.org/10.1111/ina.12806

Lu, C.-W., Liu, X.-F., Jia, Z.-F., 2020. 2019-nCoV transmission through the ocular surface must not be ignored. Lancet Lond. Engl. 395, e39–e39. https://doi.org/10.1016/S0140-6736(20)30313-5

Massarotti, N., Arpino, F., Lewis, R.W., Nithiarasu, P., 2006. Explicit and semi-implicit CBS procedures for incompressible viscous flows. Int. J. Numer. Methods Eng. 66, 1618–1640. https://doi.org/10.1002/nme.1700

Mikszewski, A., Stabile, L., Buonanno, G., Morawska, L., 2021. THE AIRBORNE CONTAGIOUSNESS OF RESPIRATORY VIRUSES: A COMPARATIVE ANALYSIS AND IMPLICATIONS FOR MITIGATION. medRxiv 2021.01.26.21250580. https://doi.org/10.1101/2021.01.26.21250580

Morawska, L., Cao, J., 2020. Airborne transmission of SARS-CoV-2: The world should face the reality. Environ. Int. 139, 105730. https://doi.org/10.1016/j.envint.2020.105730

Morawska, L., Johnson, G.R., Ristovski, Z.D., Hargreaves, M., Mengersen, K., Corbett, S., Chao, C.Y.H., Li, Y., Katoshevski, D., 2009. Size distribution and sites of origin of droplets expelled from the human respiratory tract during expiratory activities. J. Aerosol Sci. 40, 256–269. https://doi.org/10.1016/j.jaerosci.2008.11.002

Pan, Y., Zang, D., Yang, P., Poon, L.M., Wang, Q., 2020. Viral load of SARS-CoV-2 in clinical samples Yang Pan Daitao Zhang Peng Yang Leo L M Poon Quanyi Wang. The Lancet.

Raffel, M., Willert, C.E., Scarano, F., Kähler, C., Wereley, S.T., Kompenhans, J., 2018. Particle image velocimetry. A practical guide, 3rd ed. Springer International Publishing.

Scungio, M., Arpino, F., Stabile, L., Buonanno, G., 2013. Numerical simulation of ultrafine particle dispersion in urban street canyons with the spalart-allmaras turbulence model. Aerosol Air Qual. Res. 13, 1423–1437. https://doi.org/10.4209/aaqr.2012.11.0306

Sorokowska, A., Sorokowski, P., Hilpert, P., Cantarero, K., Frackowiak, T., Ahmadi, K., Alghraibeh, A.M., Aryeetey, R., Bertoni, A., Bettache, K., Blumen, S., Błażejewska, M., Bortolini, T., Butovskaya, M., Castro, F.N., Cetinkaya, H., Cunha, D., David, D., David, O.A., Dileym, F.A., Domínguez Espinosa, A. del C., Donato, S., Dronova, D., Dural, S., Fialová, J., Fisher, M., Gulbetekin, E., Hamamcıoğlu Akkaya, A., Hromatko, I., Iafrate, R., Iesyp, M., James, B., Jaranovic, J., Jiang, F., Kimamo, C.O., Kjelvik, G., Koç, F., Laar, A., de Araújo Lopes, F., Macbeth, G., Marcano, N.M., Martinez, R., Mesko, N., Molodovskaya, N., Moradi, K., Motahari, Z., Mühlhauser, A., Natividade, J.C., Ntayi, J., Oberzaucher, E., Ojedokun, O., Omar-Fauzee, M.S.B., Onyishi, I.E., Paluszak, A., Portugal, A., Razumiejczyk, E., Realo, A., Relvas, A.P., Rivas, M., Rizwan, M., Salkičević, S., Sarmány-Schuller, I., Schmehl, S., Senyk, O., Sinding, C., Stamkou, E., Stoyanova, S., Šukolová, D., Sutresna, N., Tadinac, M., Teras, A., Tinoco Ponciano, E.L., Tripathi, R., Tripathi, N., Tripathi, M., Uhryn, O., Yamamoto, M.E., Yoo, G., Pierce, J.D., 2017. Preferred Interpersonal Distances: A Global Comparison. J. Cross-Cult. Psychol. 48, 577–592. https://doi.org/10.1177/0022022117698039

Stabile, L., Massimo, A., Canale, L., Russi, A., Andrade, A., Dell'Isola, M., 2019. The Effect of Ventilation Strategies on Indoor Air Quality and Energy Consumptions in Classrooms. Buildings 9. https://doi.org/10.3390/buildings9050110

Sze To, G.N., Chao, C.Y.H., 2010. Review and comparison between the Wells–Riley and dose-response approaches to risk assessment of infectious respiratory diseases. Indoor Air 20, 2–16. https://doi.org/10.1111/j.1600-0668.2009.00621.x

Watanabe, T., Bartrand, T.A., Weir, M.H., Omura, T., Haas, C.N., 2010. Development of a dose-response model for SARS coronavirus. Risk Anal. Off. Publ. Soc. Risk Anal. 30, 1129–1138. https://doi.org/10.1111/j.1539-6924.2010.01427.x

Wells, W.F., 1934. On airborne infection: study II. Droplets and Droplet nuclei. Am. J. Epidemiol. 20, 611–618. https://doi.org/10.1093/oxfordjournals.aje.a118097

Wölfel, R., Corman, V.M., Guggemos, W., Seilmaier, M., Zange, S., Müller, M.A., Niemeyer, D., Jones, T.C., Vollmar, P., Rothe, C., Hoelscher, M., Bleicker, T., Brünink, S., Schneider, J., Ehmann, R., Zwirglmaier, K., Drosten, C., Wendtner, C., 2020. Virological assessment of hospitalized patients with COVID-2019. Nature 581, 465–469. https://doi.org/10.1038/s41586-020-2196-x

Xie, X., Li, Y., Chwang, A.T.Y., Ho, P.L., Seto, W.H., 2007. How far droplets can move in indoor environments--revisiting the Wells evaporation-falling curve. Indoor Air 17, 211–225. https://doi.org/10.1111/j.1600-0668.2007.00469.x

Zhang, N., Chen, W., Chan, P.-T., Yen, H.-L., Tang, J.W.-T., Li, Y., 2020a. Close contact behavior in indoor environment and transmission of respiratory infection. Indoor Air 30, 645–661. https://doi.org/10.1111/ina.12673

Zhang, N., Su, B., Chan, P.-T., Miao, T., Wang, P., Li, Y., 2020b. Infection Spread and High-Resolution Detection of Close Contact Behaviors. Int. J. Environ. Res. Public. Health 17. https://doi.org/10.3390/ijerph17041445